\documentclass[aps,amsmath,amssymb,superscriptaddress,prl]{revtex4-1}   
\usepackage[latin2]{inputenc}
\usepackage[T1]{fontenc}
\usepackage{amsmath,amssymb}
\usepackage{amsfonts}
\usepackage{bm}
\usepackage{dcolumn}
\usepackage{setspace}
\usepackage{graphicx}
\usepackage{color}
\usepackage{multirow}
\usepackage{verbatim}
\usepackage{epstopdf}

\def\bar{\begin{array}}
\def\ear{\end{array}}
\def\LB{\left(}
\def\RB{\right)}

\def\s{\sigma}
\def\f{\frac}

\def\r{\mathbf{r}}

\def\nn{\nonumber}

\def\m{\omega^*}

\def\p{\partial}

\def\p{\partial}

\def\l{\lambda}

\def\m{\mu}
\def\n{\nu}
\def\r{\rho}
\def\s{\sigma}
\def\t{\tau}
\def\z{\zeta}
\def\e{\epsilon}

\begin{document}

\title{Adiabatic perturbation theory for two-component systems with one heavy component}

\author{Ryan Requist} 
\affiliation{Ru\dj er Bo\v{s}kovi\'{c} Institute, Bijeni\v{c}ka cesta 54, 10000 Zagreb, Croatia}
\affiliation{Fritz Haber Center for Molecular Dynamics, Institute of Chemistry, The Hebrew University of Jerusalem, Jerusalem 91904 Israel}

\date{\today}

\begin{abstract}

Perturbation theory with respect to the kinetic energy of the heavy component of a two-component quantum system is introduced.  An effective Hamiltonian that is accurate to second order in the inverse heavy mass is derived.  It contains a new form of kinetic energy operator with a Hermitian mass tensor and a complex-valued vector potential.  All of the potentials in the effective Hamiltonian can be expressed in terms of covariant derivatives and a resolvent operator.  The most salient application of the theory is to systems of electrons and nuclei.  The accuracy of the theory is verified numerically in a model diatomic molecule and analytically in a vibronic coupling model. 

\end{abstract}

\maketitle

Perturbation theory with respect to a potential is well established, providing successively higher order approximations to an eigenfunction of the Schr\"odinger equation.  In a two-component system with one component that is much heavier than the other, it makes sense to seek a perturbation theory with respect to the kinetic energy operator of the heavy component.  The two components could be different species of particles, such as electrons and nuclei, or two different types of degrees of freedom if one type is effectively heavier than the other.  However, such a perturbation theory must be fundamentally different from standard quantum perturbation theory because the heavy component kinetic energy operator is a singular perturbation of the Schr\"odinger equation in the limit that the mass of the heavy component is taken to infinity.  

For systems of electrons and nuclei, Born and Oppenheimer (BO) \cite{born1927} and Slater \cite{slater1927} proposed an adiabatic approximation in which, at each point $x$ in the space of nuclear coordinates, the electronic state is taken to be an eigenstate, for instance the $n$th eigenstate $|n\rangle$, of the Hamiltonian with the nuclear coordinates held fixed.  The eigenstate $|n\rangle$ has a parametric dependence on $x$, $|n\rangle=|n(x)\rangle$, and the corresponding eigenenergy $E_n(x)$ is to be used as a potential in an effective Schr\"odinger equation for the wave function $\z(x)$ of the nuclei.  The product state $\psi(r,x) = \langle r |n(x)\rangle \z(x)$ with electronic coordinates $r=\{\mathbf{r}_1,\mathbf{r}_2,\ldots\}$ provides a lowest-order adiabatic approximation, called the BO approximation, which underpins most calculations in condensed matter physics.  However, unless two potentials -- the Mead-Truhlar vector potential \cite{mead1979} and a geometric scalar potential \cite{berry1989} -- are added to the effective nuclear Schr\"odinger equation, the total energy is {\it not} accurate to order $M^{-1}$, where $M$ is the mass scale of the nuclei.  These two potentials are not usually included, which means that calculations that build on the BO approximation must include order $M^{-1}$ corrections to compensate for using a less accurate starting point.  In a wide range of problems in solids 
\cite{lazzeri2006,bock2006,ferrari2007,pisana2007,piscanec2007,calandra2007,caudal2007,saitta2008,basko2009,defillipis2010,dean2010,calandra2010,gonze2011,cannuccia2012,marini2015,ponce2015,antonius2015,klimin2016,ponosov2016,dastuto2016,gali2016,nery2016,ponosov2017,allen2017,giustino2017,long2017,zhou2017,caruso2017,nery2018,marini2018,novko2018,requist2019b,miglio2020,pellegrini2022,talantsev2023,novko2023,brousseau-couture2023} and molecules \cite{makhov2014,min2015,requist2016a,li2018,crespo-otero2018,loncaric2019,agostini2019,martinazzo2022b,villaseco-arribas2022,kocak2023,manfredi2023,gardner2023,bombin2023,athavale2023}, it is important to go beyond the BO approximation.

An inherent feature of Born and Oppenheimer's original proposal \cite{born1927} is the expansion of the potential $E_n(x)$ in powers of $(m/M)^{1/4}(\mathbf{x}_i-\mathbf{x}_{i0})$ with respect to a single absolute minimum.  This is a limitation because such an approach is not universally applicable, e.g.~it does not apply to systems with multiple degenerate minima and it is impractical for arbitrary excited states.  It should be possible to formulate a systematic perturbation theory in the adiabatic limit $M\rightarrow \infty$ without expanding the potential $E_n(x)$.  The first step in this direction was Littlejohn and Weigert's (LW) pioneering development of an iterative procedure, utilizing a sequence of near-identity transformations to diagonalize a molecular or coupled-wave Hamiltonian to successively higher powers of $M^{-1/2}$ in the adiabatic limit under the condition that the nuclear kinetic energy remains constant \cite{littlejohn1993,weigert1993}.

In this paper, we introduce a singular perturbation theory with respect to the kinetic energy operator of the heavy component of a two-component system that is based solely on the smallness of $M^{-1}$ in the $M\rightarrow \infty$ limit.  Unlike previous approaches, it neither assumes an expansion of $E_n(x)$ with respect to a single reference point nor a condition on the nuclear kinetic energy.  It is applicable to solids as well as molecules.  We derive a sequence of increasingly accurate (in powers of $M^{-1}$) effective Hamiltonians in which all of the potentials are expressed in terms of a resolvent operator and covariant derivatives with respect to the coordinates of the heavy component.  

The coordinate-invariant nonrelativistic Hamiltonian of a two-component system can always be written as
\begin{align}
\mathcal{H} &= -\f{\e}{2} \f{1}{\sqrt{g_0}} \p_{\m} \big( \sqrt{g_0} g_0^{\m\n} \p_{\n} \big)  + H(x) {,}
\label{eq:H}
\end{align}
where $x = \{x^{\m}\}$ denotes the generalized, possibly curvilinear, coordinates of the heavy particles. The first term is the adiabatic Hamiltonian $H(x)$, which includes the kinetic energy of the light component and all potentials.  The second term is the heavy component kinetic energy containing the Laplace-Beltrami operator expressed in terms of an $x$-dependent metric $g_{0\m\n}$, $g_0=\mathrm{det}(g_{0\m\n})$, and $\p_{\mu} = \p/\p x^{\m}$; we use the summation convention.  
The small parameter $\e$ is 
\begin{align}
\e = \f{\f{\hbar^2}{ML^2}}{\Delta} {,}
\label{eq:epsilon}
\end{align}
where $\hbar^2/ML^2$ is an energy scale for the heavy kinetic energy and $\Delta$ is an energy scale for the light system, e.g.~a characteristic energy gap of $H$; $L$ is a characteristic length and $M$ is a characteristic mass.  This definition is useful when the light system is modeled by a Hamiltonian that does not explicitly contain a light mass $m$.  Eq.~(\ref{eq:epsilon}) includes the definition $\e=m/M$ is a special case.

The eigenvalue equation 
\begin{align}
H|n\rangle = E_n |n\rangle
\label{eq:HE}
\end{align}
defines an $x$-dependent adiabatic basis $\{|n\rangle\}$.  We write an eigenstate $|\psi\rangle$ of the two-component system in the adiabatic basis as $|\psi\rangle = \sum_n |n\rangle \z_n$.  Our main result is the derivation of a second-order effective Hamiltonian
\begin{align}
\mathcal{H}_{nn}^{\rm eff(2)} &= -\f{\e}{2} \f{1}{\sqrt{g_0}} \big( \p_{\m} - i A^{(2)*}_{\m} \big) \sqrt{g_0} h^{(2)\m\n} \big( \p_{\n} - i A^{(2)}_{\n} \big) + E_n + \e V_{geo} + \e^2 V_{2,geo} {}
\label{eq:Heff2}
\end{align} 
for a two-component system whose light component predominantly occupies the state $|n\rangle$. $h^{(2)\m\n}$ is the inverse of $h^{(2)}_{\m\n}$, a Hermitian effective mass tensor, $A_{\n}^{(2)}$ is a complex-valued effective gauge potential, and $V_{geo}$ and $V_{2,geo}$ are effective potentials.  The formulas for all quantities are given below. 

To derive $\mathcal{H}_{nn}^{\rm eff(2)}$, and more generally a $p$th-order effective Hamiltonian $\mathcal{H}_{nn}^{\rm eff(p)}$, we perform a sequence of near-identity unitary transformations 
\begin{align}
|\psi\rangle &= U_1 U_2 \ldots U_p |\psi^{(p)}\rangle 
\end{align}
to a representation in which the $n$th eigenstate is decoupled from all others to order $\e^p$, i.e.
\begin{align}
\mathcal{H}^{(p)}_{ln} f &= \langle l | U_p^{\dag} \ldots U_2^{\dag} U_1^{\dag} \mathcal{H} U_1 U_2 \ldots U_p \big( |n\rangle f \big) \nn \\
&= \mathcal{O}(\e^{p+\f{1}{2}}) 
\label{eq:condition:p}
\end{align}
for $l\neq n$; $f$ is a test function which is allowed to be $\e$ dependent such that $\p_{\m} f/f = \mathcal{O}(\e^{-1/2})$, but it cannot have a stronger divergence than this in the $\e\rightarrow 0$ limit.  The unitary transformations 
\begin{align}
U_k = e^{i\e^k G_k(x,p)} 
\end{align}
become successively closer to the identity.  $U_k$ operates on both light and heavy variables.
If the light-component Hilbert space is truncated to a finite-dimensional subspace, the action of $U_k$ in that space can be represented by matrix multiplication.  Simultaneously, $U_k$ acts as a differential operator on functions of the coordinates of the heavy component.  Starting at first order and expanding the transformed Hamiltonian $U_1^{\dag} \mathcal{H} U_1$ in powers of $\e$, we obtain the first-order effective Hamiltonian
\begin{align}
\mathcal{H}^{\rm eff(1)} = H + \e T - i\e [G_1,H] {,}
\end{align}
where $T$ is the operator multiplied by $\e$ in Eq.~(\ref{eq:H}).
The condition in Eq.~(\ref{eq:condition:p}) will be satisfied if $G_1$ satisfies
\begin{align}
\langle l | [G_1,H] \big( |n\rangle f \big) = -i \langle l | T \big( |n\rangle f \big) \quad \mathrm{for}\;\mathrm{all} \; l\neq n {}
\label{eq:condition:G1}
\end{align}
for the same type of test function $f$ used in Eq.~(\ref{eq:condition:p}).
Equation~(\ref{eq:condition:G1}) does not determine the diagonal elements of $G_1$, which we choose to be zero.  Once we have found a suitable $G_1$, the coupling to other states will be order $\epsilon^{3/2}$.  Since coupling of this magnitude is negligible when we are working to first order in $\e$, we write the wave function in this representation as $|\psi_n^{(1)}\rangle = |n\rangle \z_n^{(1)}$, neglecting all states of the light system except $|n\rangle$.  Substituting $|\psi_n^{(1)}\rangle$ into the effective eigenvalue equation $\mathcal{H}^{\rm eff(1)} |\psi^{(1)}\rangle = \lambda^{(1)} |\psi^{(1)}\rangle$ and projecting onto $|n\rangle$, gives the first-order effective Schr\"odinger equation
\begin{align}
-\f{\e}{2} \f{1}{\sqrt{g_0}} \big( \p_{\m} - i A_{\m} \big) \sqrt{g_0} g_0^{\m\n} \big( \p_{\n} - i A_{\n} \big) \z_{nm}^{(1)} + E_n \z_{nm}^{(1)} + \e V_{geo} \z_{nm}^{(1)} = \lambda_{nm}^{(1)} \z_{nm}^{(1)}& {,}
\label{eq:1st}
\end{align}
where $A_{\n} = i \langle n | \p_{\n} n\rangle$ is the Mead-Truhlar gauge potential \cite{mead1979,mead1980a}, $V_{geo}$ is the geometric scalar potential \cite{berry1989}
\begin{align}
V_{geo} &= \f{1}{2} g_0^{\m\n} \sum_{l\neq n} A_{\m,nl} A_{\n,ln} \nn \\
&= \f{1}{2} g_0^{\m\n} \mathrm{Re} \langle D_{\m} n | D_{\n} n\rangle  {,}
\end{align}
and $D_{\m}=\p_{\m}+i A_{\m}$ is the gauge-covariant derivative acting on $|n\rangle$.
The eigenvalue $\lambda_{nm}^{(1)}$ in Eq.~(\ref{eq:1st}) approximates the total eigenenergy of the two-component system to order $\e$.  Equation (\ref{eq:1st}) has been written in a gauge- and coordinate-invariant form and all of the potentials it contains have been expressed in terms of $|n\rangle$ and its covariant derivatives.  In particular, sums over adiabatic eigenstates have been eliminated, so there are no off-diagonal quantities such as the nonadiabatic coupling $A_{\n,ln}=i\langle l |\p_{\n} n\rangle$. 
The potentials have geometric significance, e.g.~the potential $V_{geo}$ was shown to be expressible in terms of the Provost-Vallee metric \cite{provost1980}, $g_{\m\n}=\mathrm{Re}\langle D_{\m} n | D_{\n} n\rangle$, by Berry \cite{berry1989,berry1990}.  In the special case of an $x$-independent metric $g_{0\m\n}$, Eq.~(\ref{eq:1st}) reproduces the non-coordinate-invariant effective equation derived previously \cite{berry1989}. 

We now proceed to second order and look for a unitary transformation $U_2=e^{i\e^2 G_2}$ that decouples the $n$th state to order $\e^2$.
The second-order effective Hamiltonian is 
\begin{align}
\mathcal{H}^{\rm eff(2)} = H + \e T - i\e [G_1,H] -i \f{\e^2}{2} [G_1,T] - i \e^2 [G_2,H]  {.}
\end{align}
To evaluate how $\mathcal{H}^{\rm eff(2)}$ operates on $|\psi_n^{(2)}\rangle = |n\rangle \z_n^{(2)}$, we need to solve for $G_1$ and evaluate the commutator $[G_1,T]$, but we do not need to know $G_2$, just as we did not need to know $G_1$ to evaluate how $\mathcal{H}^{\rm eff(1)}$ acts on $|\psi_n^{(1)}\rangle$.  Since $T$ is a second-order differential operator, we look for a solution to Eq.~(\ref{eq:condition:G1}) of the form
\begin{align}
G_{1,ln} = - \f{1}{\sqrt{g_0}} \sum_m &(\p_{\m} \delta_{lm} - iA_{\m,lm}) \sqrt{W_{ln}} W_{ln}^{\m\n} (\p_{\n}\delta_{mn} - iA_{\n,mn}) + U_{ln} {,}
\end{align}
where $W_{ln}=\det(W_{\m\n,ln})$.  $G_{1}$ acts as a Hermitian operator on the vector $(\z_1,\z_2,\ldots)^T$ if $W_{ln}^{\m\n}=W_{nl}^{\n\m *}$ and $U_{ln}=U_{nl}^*$.  $G_1$ will satisfy Eq.~(\ref{eq:condition:G1}) if we choose
\begin{align}
\sqrt{W_{ln}} W_{ln}^{\m\n} = \f{i}{2} (E_l-E_n)^{-1} \sqrt{g_0} g_0^{\m\n} 
\end{align}
and
\begin{align}
U_{ln} = -\f{1}{2} (E_l-E_n)^{-2} g_0^{\m\n} \p_{\m}(E_l+E_n) A_{\n,ln} {.}
\end{align}
Littlejohn and Weigert \cite{littlejohn1993,weigert1993} and Refs.~\cite{teufel2003} and \cite{matyus2019} obtained a different $G_1$ and hence different effective Hamiltonians because they hold the nuclear kinetic energy constant while taking the $M\rightarrow\infty$ limit (see Supplemental Material \cite{suppl} for a comparison with the present theory).

The second-order effective Schr\"odinger equation 
\begin{align}
-\f{\e}{2} \f{1}{\sqrt{g_0}} \big( \p_{\m} - i A^{(2)*}_{\m} \big) \sqrt{g_0} h^{(2)\m\n} \big( \p_{\n} - i A^{(2)}_{\n} \big) \z_{nm}^{(2)} 
+ (E_{nm} + \e V_{geo} + \e^2 V_{2,geo}) \z_{nm}^{(2)} = \lambda_{nm}^{(2)} \z_{nm}^{(2)} & {,}
\label{eq:2nd}
\end{align} 
which contains the second-order Hamiltonian $\mathcal{H}_{nn}^{\rm eff(2)}$, can be derived by making the energy 
\begin{align}
\lambda_n^{(2)} &=\f{\e}{2} \int \sqrt{g_0} (\p_{\n}+iA_{\n})\z_n^{(2)*} g_0^{\n\s} (\p_{\s}-iA_{\s}) \z_n^{(2)} dx + \int \sqrt{g_0} (E_n+\e V_{geo}) |\z_n^{(2)}|^2 dx +\e^2 \lambda_{2n}
\end{align}
stationary with respect to variations of $\z_n^{(2)}$.  The second-order contribution is 
\begin{align}
\lambda_{2n} &= -\f{i}{2} \int \sqrt{g_0} G_1 \big( \z_n^{(2)*} \langle n | \big) T \big( |n\rangle \z_n^{(2)} \big) dx +\f{i}{2} \int \sqrt{g_0} T \big( \z_n^{(2)*} \langle n | \big) G_1 \big( |n\rangle \z_n^{(2)} \big) dx {.}
\end{align}
In the Euler-Lagrange equation, $\lambda_{2n}$ generates on the order of 225 terms containing inner products such as $\langle l | \p_{\m} H | m\rangle$.  Remarkably, all of these terms can be absorbed into the complex-valued gauge potential $A_{\m}^{(2)} = A_{\m} + \e A_{2\m}$, the effective mass tensor $h^{(2)}_{\m\n}=g_{0\m\n} + \e h_{2\m\n}$, and the effective potential $V_{2,geo}$, which are given by the following compact formulas:
\begin{align}
A_{2\n} &= \langle D_{\n} n |(E_n-H)^{-1} |\nabla^2 n\rangle - \langle D_{\n} n |(E_n-H)^{-2}| D_{\r} n\rangle g_0^{\r\s} \p_{\s} E_n 
\end{align}
\begin{align}
h_{2\m\n} = -2 \langle D_{\m} n | (E_n-H)^{-1} | D_{\n} n\rangle
\end{align}
\begin{align}
V_{2,geo} &= \f{1}{4} \langle \nabla^2 n | (E_n-H)^{-1} | \nabla^2 n \rangle -\f{1}{2}  g_0^{\m\n} \p_{\n} E_n \mathrm{Re} \langle D_{\m} n |(E_n-H)^{-2} | \nabla^2 n \rangle {}
\end{align}
with
\begin{align}
|\nabla^2 n\rangle &\equiv g_0^{\r\s} (1-|n\rangle\langle n|) |D_{\r} D_{\s} n\rangle - g_0^{\r\s} \Gamma_{0\r\s}^{\t} |D_{\t} n\rangle 
\end{align}
where $\Gamma_{0\m\n}^{\l}$ is the Christoffel symbol associated with $g_{0\m\n}$.

Equation (\ref{eq:2nd}) is gauge- and coordinate-invariant and contains a new form of Hermitian kinetic energy operator.  The quantum number $n$ labels the state of the light component in accordance with Eq.~(\ref{eq:HE}), while $m$ labels the heavy component.  All of the potentials have been written in terms of the resolvent $(E_n-H)^{-1}$, $E_n$, $|n\rangle$, and covariant derivatives of $|n\rangle$, i.e.~sums over adiabatic eigenstates have been eliminated, as in the first-order equation.  This is relevant to the development of efficient approaches for evaluating the potentials that do not require explicit summations over adiabatic eigenstates, such as the Chebyshev expansion algorithm or kernel polynomial methods \cite{kosloff1994,weisse2006}. The second-order correction $A_{2\n}$ is manifestly gauge invariant.  Hence the quantities $A_{\n}^{(2)}$ and $A_{\n}^{(2)*}$ that appear in Eq.~(\ref{eq:2nd}) transform in exactly the same way as $A_{\n}$ under the gauge transformation $|n\rangle \rightarrow e^{i\l} |n\rangle$. The effective kinetic energy operator in Eq.~(\ref{eq:2nd}) can be expressed in several equivalent forms, e.g.~$-(\e/2) h^{(2)\m\n} \hat{\nabla}_{\m}^{(2)} \hat{\nabla}_{\n}^{(2)} = -(\e/2) h^{(2)\m\n} D_{\m} D_{\n} + (\e/2) h^{(2)\m\n} \Lambda_{\m\n}^{(2)\l} D_{\l} + V^{\prime}$, where $\hat{\nabla}^{(2)}$ is a quantum covariant derivative with the symbol $\Lambda_{\m\n}^{(2)\l}$, similar to the one introduced in Ref.~\cite{requist2023a}.  Details of the derivation of Eq.~(\ref{eq:2nd}) will be presented in a separate publication.  

The procedure outlined above can be carried out to arbitrary order, and at each order the potentials can be expressed in terms of the resolvent, $E_n$, $|n\rangle$, and its covariant derivatives.  This is analogous to what was found in the adiabatic perturbation theory for explicitly time dependent Hamiltonians developed in Ref.~\cite{requist2023a}, which is a singular perturbation theory in the differential operator $i d/dt$ instead of the kinetic energy operator.

The eigenvalue $\lambda_{nm}^{(p)}$ is an order $\e^p$ approximation to the eigenenergy of the two-component system, i.e.~the error scales as a higher power of $\e$ than $\e^p$.  Assuming that an approximate solution $\z_{nm}^{(p)}$ of the $p$th-order effective Schr\"odinger equation can be obtained, our theory also produces an approximation to the corresponding eigenstate of the two-component system, which can be used to evaluate any observable.  To obtain the wave function $|\psi\rangle$ of the two-component system in the original representation, we operate with $U_1\ldots U_p$ on $|\psi^{(p)}\rangle$.  In numerical calculations, it is convenient to introduce a complete basis $\{|b\rangle\}$ for the $x$-space and write e.g.
\begin{align}
|\psi^{(1)}\rangle &= \sum_{b} |n\rangle |b\rangle c_{nb}^{(1)} {,}
\end{align}
where $c_{nb}^{(1)}$ are constant coefficients.  Then, we have
\begin{align}
|\psi\rangle &= U_1 |\psi^{(1)}\rangle \nn \\
&= \sum_b |n\rangle |b\rangle c_{nb}^{(1)} + i\e \sum_{lab} |m\rangle |a\rangle G_{1,la,nb} c_{nb}^{(1)} + \mathcal{O}(\e^2) {}
\label{eq:psi:from:psi1}
\end{align}
with $G_{1,la,nb} = \langle l| \langle a| G_1 |n\rangle |b\rangle$.

\begin{figure}[t]
\begin{center}
\includegraphics[width=0.45\columnwidth]{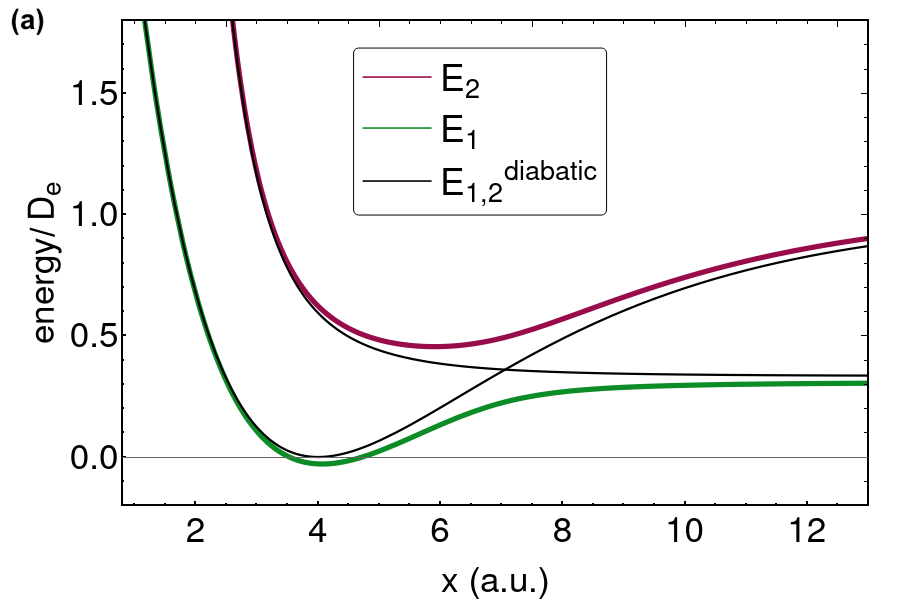} 
\includegraphics[width=0.45\columnwidth]{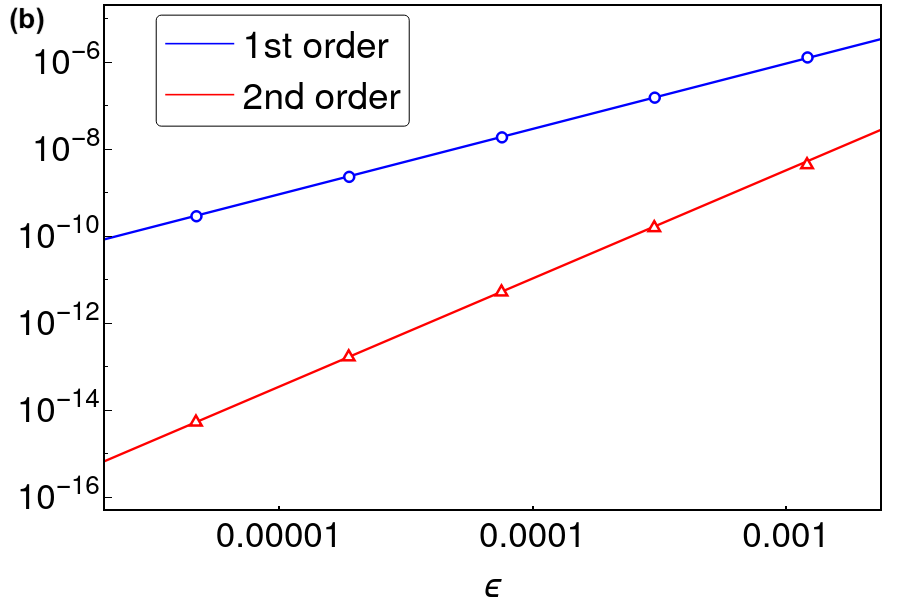} 
\end{center}
\caption{Morse diatomic molecule.  (a) Adiabatic and diabatic potential energy surfaces versus $x$.  (b) Log-log plot of the error in the ground state energy in the first-order (blue circles) and second-order (red triangles) APT approximations.  Estimated asymptotic linear behaviors are (b) $-3.49 + 1.50 \ln \e$ and $-2.33 + 2.49 \ln \e$ for first- and second-order approximations, respectively. \label{fig:morse}}
\end{figure}

Now we exemplify adiabatic perturbation theory in two specific systems.  
The first is a model two-level diatomic molecule with the Hamiltonian
\begin{align}
\mathcal{H} &= -\f{\hbar^2}{2M} \f{d^2}{dx^2} + \LB \bar{cc} D_e (1- e^{-a (x-x_0)})^2 & -c \\ -c & D_e-\Delta+\f{b}{x^4} \ear \RB  {,}
\label{eq:morse}
\end{align}
where the $(1,1)$ element of the matrix is the Morse potential.  In Fig.~\ref{fig:morse}a, the adiabatic energies $E_1(x)$ and $E_2(x)$ and diabatic energies [the (1,1) and (2,2) elements of the matrix] are plotted for $M=10^4 m_e$, $D_e=0.015$ E$_h$, $\Delta=0.010$ E$_h$, $x_0=4.0 a_0$, $a=0.30$ $a_0^{-1}$, $b=1.00 a_0^4$, and $c=0.002$ E$_h$.

The metric in this coordinate system is trivial, $g_{0\m\n}=\delta_{\m\n}$, yet an important property of the first and second-order effective Schr\"odinger equations in Eqs.~(\ref{eq:1st}) and (\ref{eq:2nd}) is their coordinate invariance.  To test the coordinate-invariant form of the equation in a case with a nontrivial $g_{0\m\n}$, we change the independent variable according to
\begin{align}
q = \sqrt{2} aL \sinh \f{x-x_0}{L} {,}
\end{align}
which changes the kinetic energy operator to the standard form in Eq.~(\ref{eq:H}) with $g_{0qq}=(1+q^2/2a^2L^2)^{-1}$ and the small parameter $\e=2\hbar^2 a^2/MD_e$. Harmonic oscillator basis functions $\phi_i$ are used to express the Schr\"odinger equation as a generalized eigenvalue problem
\begin{align}
\sum_j \mathcal{H}_{ij} \psi_{nj} = \lambda_n \sum_j S_{ij} \psi_{nj} {,}
\end{align}
where $\mathcal{H}_{ij} = \int \sqrt{g_{0}} \phi_i^* \mathcal{H} \phi_j dq$ and $S_{ij} = \int \sqrt{g_{0}} \phi_i^* \phi_j dq$.

Figure \ref{fig:morse}b shows that the error in the eigenenergy $\lambda_{00}^{(1)}$ is order $\e^{3/2}$, while the error in $\lambda_{00}^{(2)}$ is order $\e^{5/2}$.  This is expected: the eigenenergy of the $p$th-order effective Schr\"odinger equation is accurate to order $\e^p$.  A numerical analysis identical to that performed in Fig.~\ref{fig:morse}b shows that the error in $\langle q\rangle$ and $\langle q\sigma_z \rangle$, where $\sigma_z$ is the Pauli matrix, is order $\e^{3/2}$ in the first-order approximation.  

To illustrate the first- and second-order effective Schr\"odinger equations in a case in which the Mead-Truhlar gauge potential is nontrivial, we now consider the linear $E\otimes e$ Jahn-Teller model.  In a basis of real-valued electronic states $\{|u\rangle,|g\rangle\}$, the Hamiltonian is
\begin{align}
\mathcal{H} = -\f{\hbar^2}{2M} \LB \f{d^2}{dx_1^2} +\f{d^2}{dx_2^2}\RB &+ \f{K}{2} \big(x_1^2+x_2^2\big) I + g \LB \bar{cc} x_1 & -x_2 \\ -x_2 & -x_1 \ear \RB {.}
\end{align}
We can achieve a separation of variables by introducing polar coordinates $x=\sqrt{x_1^2+x_2^2}$, $\phi=\mathrm{arctan2}(x_1,x_2)$, transforming to a basis $|\pm\rangle = \f{1}{\sqrt{2}}(|g\rangle \pm i |u\rangle)$ of current-carrying states, and considering a state with angular momentum quantum number $j$:
\begin{align}
|\psi_{\n j}\rangle = \f{1}{\sqrt{2\pi}} \LB \bar{c} a_{\n j} e^{i(j-1/2)\phi} \\ b_{\n j} e^{i(j+1/2)\phi} \ear \RB {.}
\end{align}
The angular momentum operator is $J = -i\p/\p \phi + \f{1}{2} \sigma_3$; its quantum number $j$ takes half-integer values \cite{moffitt1957a,moffitt1957b,longuet-higgins1958}. 

To put the Schr\"odinger equation in the standard form of Eq.~(\ref{eq:H}), we change the radial variable to $q=x/x_0$, where $x_0=g/K$ is the minimum of the radial potential $Kx^2/2-gx$, use the energy unit $2E_{JT}=g^2/K$, and define $\e = \hbar^2 K^3/Mg^4$.  The Mead-Truhlar gauge potential is $A_{\phi}=-j$ and the geometric potential is
\begin{align}
V_{geo} = \f{1}{8q^2} {.}
\end{align}
This fixes all of the potentials in the first-order equation, which for the ground state $n=0$ takes the form
\begin{align}
-\f{\e}{2} \bigg( \f{d^2}{dq^2} &+ \f{1}{q} \f{d}{dq} \bigg) \z_{0\n}^{(1)} + \f{\e A_{\phi}^2}{2q^2} \z_{0\n}^{(1)}+\bigg( \f{1}{2} q^2 - q + \e V_{geo} \bigg) \z_{0\n}^{(1)} = \l_{0\n}^{(1)} \z_{0\n}^{(1)} {}
\end{align}  
with the centrifugal potential $\e A_{\phi}^2/2q^2 = \e j^2/2q^2$. 

To define the second-order equation, we need to know the effective mass tensor $h^{(2)}_{\m\n}$, the geometric potential $V_{2,geo}$, and the gauge potential $A_{2\n}$. The only nonzero element of $h_{2\m\n}$ is $h_{2\phi\phi} = 1/(4q)$.  The kinetic energy therefore contains the additional term
\begin{align}
-\f{\e^2}{2} A_{\phi} g_0^{\phi\phi} h_{2\phi\phi} g_0^{\phi\phi} A_{\phi}  = -\e^2 \f{A_{\phi}^2}{8q^5} {,}
\label{eq:extra}
\end{align}
which acts as a perturbation of the centrifugal potential corresponding to a change of the moment of inertia from $I_0=q^2$ to $I^{(2)}=I_0+\e h_{2\phi\phi}$. Since $A_{2\n}$ and $V_{2,geo}$ are zero, we have determined all of the potentials in the second-order equation, and our analytical calculations verify that it reproduces the results of O'Brien and Pooler \cite{obrien1979}.  For example, the second-order eigenenergy 
\begin{align}
\f{\lambda_{0\n}^{(2)}}{\hbar\omega} &= -\e^{-1/2} \f{1}{2} + \n + \f{1}{2} + \e^{1/2} \f{j^2}{2} + \e \f{3j^2}{2} (\n+1/2) + \e^{3/2} \f{j^2}{8} [30(\n^2+\n+1/2)-1] - \e^{3/2} \f{j^4}{2}  \nn \\
&\quad+ \e^2 \f{5j^2}{16} [28(\n+1/2)^3+29(\n+1/2)] - \e^2 \f{57j^4}{8} (\n+1/2)  
\label{eq:lambda:JT}
\end{align}
is accurate to order $\e^2$ as expected.  The eigenvalue of the first-order equation is accurate to order $\e$. For $j=1/2$, the potential in Eq.~(\ref{eq:extra}) has been obtained \cite{requist2017} from an asymptotic analysis of the nonlinear equations of the exact factorization method \cite{gidopoulos2014,abedi2010,hunter1975}.  Unlike the asymptotic analyses in Refs.~\cite{obrien1979} and \cite{requist2017}, which were tailored to the ground states of a specific class of symmetric Jahn-Teller problems, the theory developed here applies to arbitrary eigenstates of any light-heavy system, provided only that the eigenstate in question occupies predominantly a single state $|n\rangle$ of the light system.

In the first- and second-order effective Hamiltonians, the momentum operator $-i\p_{\m}$ only appears through the gauge-covariant derivative $D_{\m}=\p_{\m}-iA_{\m}$.  In the perturbation theory developed here, this structure is preserved at every order, and the $A_{\m}$-containing terms are not assumed to be higher order in $\e$ than the $\p_{\m}$-containing terms.  In the $E\otimes e$ Jahn-Teller model, the $A_{\m}$-containing term in Eq.~(\ref{eq:1st}), $\e A_{\phi}^2/(2q^2)$, leads to the energy correction $\e^{1/2} (j^2/2) \hbar \omega = \e j^2 E_{JT}$ in Eq.~(\ref{eq:lambda:JT}).  For $j$ of order 1, this correction is order $\e^{1/2}$ smaller than the radial vibrational energy $(\n+1/2)\hbar\omega$.  However, for high angular momentum states with $j$ of order $\e^{-1/4}$, the energy correction from $\e A_{\phi}^2/(2q^2)$ is of the same order as the radial vibrational energy, demonstrating that there are states for which $A_{\m} \z_{n}$ should not be considered to be of higher order in $\epsilon$ than $\p_{\m} \z_{n}$.
The appearance of the Mead-Truhlar vector potential at every order in our hierarchy of approximations, particularly in the generators $G_k$, is important for perturbative nonadiabatic 
descriptions of vibrational modes in pseudorotating molecules and crystal defects \cite{kendrick1997,perebeinos2005,allen2005,requist2016b,ryabinkin2017,kato2022,lyakhov2023} and phonons with angular momentum \cite{zhang2014,zhu2018,ishito2023,ueda2023}.

\section{Conclusions}

We have introduced an adiabatic perturbation theory for two-component systems that yields a sequence of effective Hamiltonians that are accurate to successively higher powers of $M^{-1}$.  It can be understood as a singular perturbation theory in the kinetic energy operator of the heavy component.  Adiabatic perturbation theory is quite different from other many-body perturbation theories, especially in electron-phonon systems, which usually involve approximations whose error in terms of $M^{-1}$ is unknown.  The BO approximation is the conventional starting point in practical calculations of solids, yet we have pointed out that its assumption of an expansion with respect to a single reference point is a serious limitation, which is avoided in adiabatic perturbation theory.

\section{acknowledgments}

I am grateful to R.~G.~Littlejohn for discussions leading to improvements in the paper.  Stimulating discussions with A.~G.~Abanov are also gratefully acknowledged.  


\bibliography{apt-static-2023}

\end{document}


%
%


\thispagestyle{empty}

\begin{center}
{\Large \textbf{Supplemental material}}
\end{center}
\vspace{0.6cm}

We report here analytical and numerical calculations comparing the perturbation theory developed in the main text with Littlejohn and Weigert's (LW) theory \cite{littlejohn1993,weigert1993} (see also Refs.~\cite{teufel2003}, \cite{matyus2019}, and \cite{littlejohn2024} for subsequent developments).
%
%
%
The perturbation theory developed in these references
%
%
is designed for large-amplitude (order 1 with respect to $M^{-1}$) nuclear motion in molecules and therefore makes assumptions about how the dynamical variables scale with $M$ in the $M\rightarrow \infty$ limit.  In particular, the nuclear kinetic energy is held constant while taking the $M\rightarrow \infty$ limit, which implies that the de Broglie wave length of the nuclear wave function becomes much shorter than the length scale of $E_n(x)$.  
For further discussion of this type of limit, see Ref.~\cite{mead2006}.

\begin{figure}[h]
\begin{center}
     \includegraphics[width=0.50\textwidth]{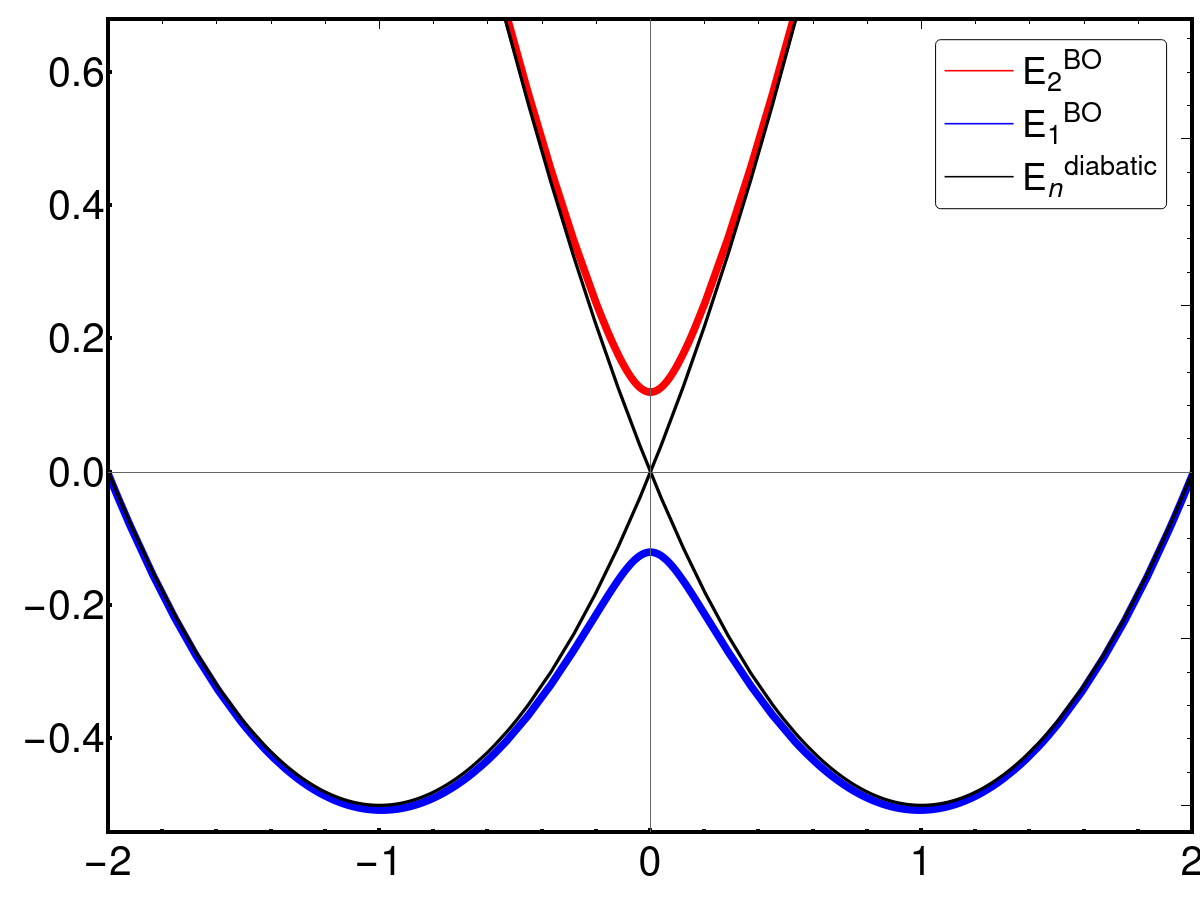} 
     \caption{Double-well model.  Adiabatic and diabatic potentials in units of $2E_{JT}$ versus $q$ for coupling parameter $\gamma=0.12$. \label{fig:doublewell}}
\end{center}
\end{figure}

We consider a diatomic molecule with the Hamiltonian 
\begin{align}
\mathcal{H} &= -\f{\hbar^2}{2M} \f{d^2}{dx^2} + \f{K}{2} x^2 I + \LB \bar{cc} g x & -B\\ -B & -g x \ear \RB {}
\label{eq:double:well}
\end{align}
in a basis of two electronic states $\{|1\rangle,|2\rangle\}$.
%
%
Changing the independent variable to $q=x/x_0$ with $x_0=g/K$, dividing by $2E_{JT}=g^2/K$, 
and  
introducing the coupling parameter $\gamma = B/(2E_{JT})$, and the dimensionless small parameter
\begin{align}
\e = \f{\hbar^2 K^3}{Mg^4} {,}
\end{align}
converts the Schr\"odinger equation to the standard form
\begin{align}
-\f{\e}{2} \f{d^2}{dq^2} |\psi\rangle + \LB \bar{cc} \f{1}{2} q^2 + q & -\gamma \\ -\gamma & \f{1}{2} q^2 - q \ear \RB |\psi\rangle = \f{\lambda}{2E_{JT}} |\psi\rangle 
{.}
\label{eq:schroedinger:standard}
\end{align}
%
The adiabatic 
%
%
eigenfunctions are parametrized as
\begin{align}
|n\rangle = \LB \bar{c} \cos\f{\theta}{2} \\[0.2cm] \sin\f{\theta}{2} \ear \RB \qquad |m\rangle = \LB \bar{c} -\sin\f{\theta}{2} \\[0.2cm] \cos\f{\theta}{2} \ear \RB {}
\end{align}
with $\theta = \mathrm{arctan2}(-q, \gamma)$.
The lower energy adiabatic potential $E_n(x)$ is a symmetric double well, as seen in Fig.~\ref{fig:doublewell}.

An accurate numerical solution is obtained by solving a generalized eigenvalue problem
in a nonorthogonal product basis $|i\rangle |\a\n\rangle$, where $|i\rangle=|1\rangle,|2\rangle$ denotes the electronic state and $|\a\n\rangle$ is a harmonic oscillator basis state associated with the left or right well ($\a=L,R$), e.g.~for the right well
\begin{align}
\langle q | R\n\rangle &= \f{1}{\sqrt{2^{\n} \n!}} \LB \f{1}{\pi \sqrt{\e}} \RB^{1/4} H_{\n}\bigg(\f{q-1}{\e^{1/4}}\bigg) e^{-\f{(q-1)^2}{2\sqrt{\e}}} {.}
\end{align}

It is instructive to list for this model the first- and second-order APT effective Schr\"odinger equations 
side-by-side with the corresponding equations in LW's theory:
\begin{align}
&-\f{\e}{2} \f{d^2}{dq^2} f_{n\n}^{(1)} + E_n f_{n\n}^{(1)} = \l_{n\n}^{LW(1)} f_{n\n}^{(1)} & & (LW1)
\label{eq:LW1} \\
&-\f{\e}{2} \f{d^2}{dq^2} \z_{n\n}^{(1)} + \big( E_n + \e V_{geo}\big) \z_{n\n}^{(1)} = \l_{n\n}^{APT(1)} \z_{n\n}^{(1)} & & (APT1)
\label{eq:APT1} \\
&-\f{\e}{2} \f{d}{dq} (1 - \e h_{2qq}) \f{d}{dq} f_{n\n}^{(2)} + \big( E_n + \e V_{geo} \big) f_{n\n}^{(2)} = \l_{n\n}^{LW(2)} f_{n\n}^{(2)} & & (LW2)
\label{eq:LW2} \\
&-\f{\e}{2} \Big( \f{d}{dq} +i \e A_{2q}^*\Big) (1 - \e h_{2qq}) \Big( \f{d}{dq} -i \e A_{2q} \Big) \z_{n\n}^{(2)} + \big( E_n + \e V_{geo} + \e^2 V_{2geo} \big) \z_{n\n}^{(2)} = \l_{n\n}^{APT(2)} \z_{n\n}^{(2)} & & (APT2) {.}
\label{eq:APT2} 
\end{align}
The lowest-order equation in LW's theory, Eq.~(\ref{eq:LW1}), is the Born-Oppenheimer equation; the only potential it contains is the adiabatic potential $E_n$.
The lowest-order APT equation is Eq.~(\ref{eq:APT1}).  It is similar to Eq.~(\ref{eq:LW1}), except it contains the geometric potential $\e V_{geo}$. 
%
LW's next-higher-order equation, Eq.~(\ref{eq:LW2}), contains the inverse effective mass tensor $h^{(2)qq} = 1 - \e h_{2qq}$.  The $h_{2qq}$ term is contained 
in the general equations derived by LW \cite{littlejohn1993,weigert1993}. 
The next-higher-order APT equation, Eq.~(\ref{eq:APT2}), is similar to Eq.~(\ref{eq:LW2}), except it contains the additional potentials $\e A_{2q}$ and $\e^2 V_{2,geo}$.  
While the Born-Oppenheimer equation is the lowest-order equation in LW's theory, it does not appear in the APT hierarchy because the lowest-order APT equation, Eq.~(\ref{eq:APT1}), is already more accurate, being, of necessity, accurate to order $\e$, while the Born-Oppenheimer equation is only accurate to $\e^{1/2}$. 


Since each one of Eqs.~(\ref{eq:LW1})-(\ref{eq:APT2}) has a different order of accuracy, it is helpful to denote the {\it order} of the approximation 
not according to its accuracy but according to the number of near-identity transformations (from the sequence $U_1,U_2, \ldots$) needed to reach a given level of approximation.  Thus, the approximation in Eq.~(\ref{eq:LW1}) is labeled LW(1), the approximation in Eq.~(\ref{eq:APT1}) is labeled APT(1), etc.




\begin{figure}[h]
\begin{center}
     \includegraphics[width=0.50\textwidth]{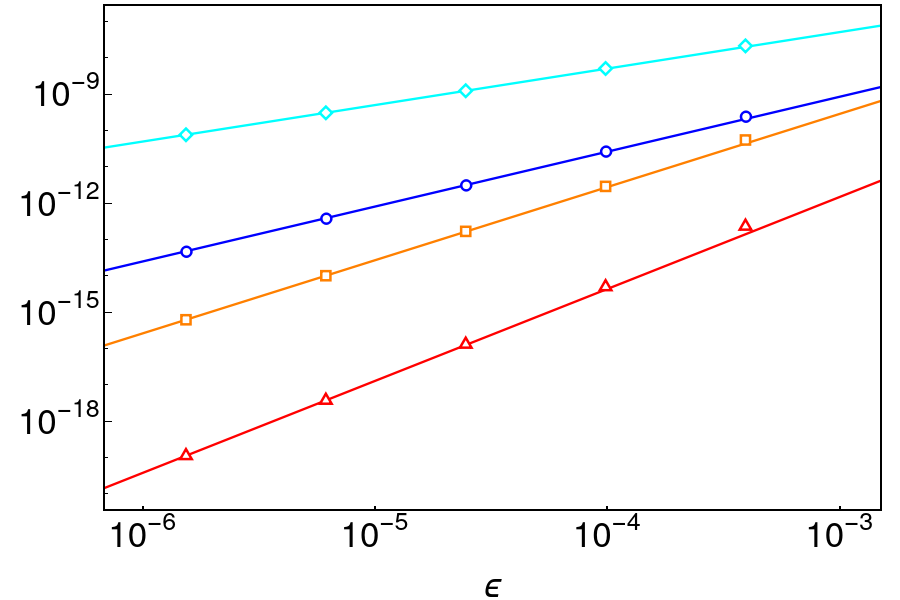} 
     \caption{Double-well model.  Log-log plots of the error in the ground state energy in APT(1) and APT(2) (blue circles and red triangles) and LW(1) and LW(2) (cyan diamonds and orange squares) approximations versus $\e$.  Estimated asymptotic linear behaviors are $-10.44 + 1.51 \ln\e$ and $-9.73 + 2.53 \ln\e$ for APT(1) and APT(2) and $-9.84 + 1.00 \ln\e$ and $-8.05 + 2.01 \ln\e$ for LW(1) and LW(2).  The coupling parameter is $\gamma=0.02$. \label{fig:gs:energy}}
\end{center}
\end{figure}



We now compare the accuracy of APT and LW's theory for  
the ground state or any sufficiently low energy excited state, holding its quantum number fixed in the $\e\rightarrow 0$ limit.
In this case, Fig.~\ref{fig:gs:energy} shows that the LW($p$) approximation is less accurate than the APT($p$) approximation. 
For example, the LW(1) energy eigenvalue $\lambda^{LW(1)}$ is accurate to order $\e^{1/2}$, i.e.~its has an error of order $\e$.
Adding the order $\e$ correction $\e V_{geo}$ to the potential energy surface of the LW(1) equation, as is done in APT(1), removes the order $\e$ error in $\lambda^{LW(1)}$, thereby increasing the accuracy of the eigenenergy to order $\e$, i.e.~the error of $\lambda^{APT(1)}$ is of order $\e^{3/2}$.
The accuracy of $\lambda^{LW(2)}$ is order $\e^{3/2}$, i.e.~it has an error of order $\e^2$. 
Adding $\e^2 V_{2,geo}$ to the potential energy surface and $\e A_{2\n}$ to the gauge potential in the LW(2) equation, as is done in APT(2), increases the accuracy of the eigenenergy to order $\e^2$, i.e.~the error of $\lambda^{APT(2)}$ is of order $\e^{5/2}$. 

\begin{figure}[h]
\begin{center}
     \includegraphics[width=0.50\textwidth]{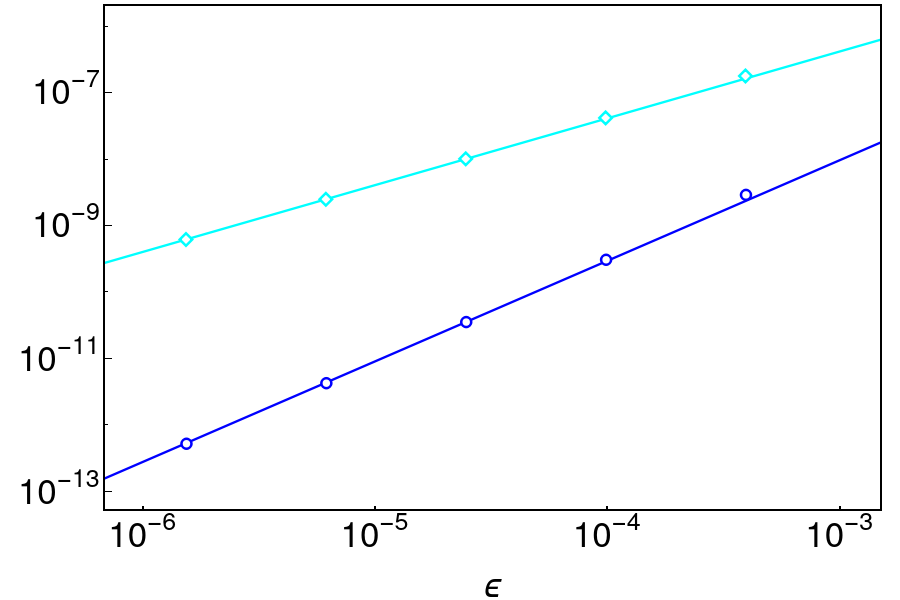} 
     \caption{Double-well model.  Log-log plots of the error in the ground state $\langle q^2 \rangle$ in APT(1) (blue circles) and LW(1) (cyan diamonds) approximations versus $\e$.  Estimated asymptotic linear behaviors are $-8.00 + 1.51 \ln\e$ for APT(1) and $-7.74 + 1.00 \ln\e$ for LW(1). The coupling parameter is $\gamma=0.02$. \label{fig:gs:qSq}}
\end{center}
\end{figure}

\begin{figure}[h]
\begin{center}
     \includegraphics[width=0.50\textwidth]{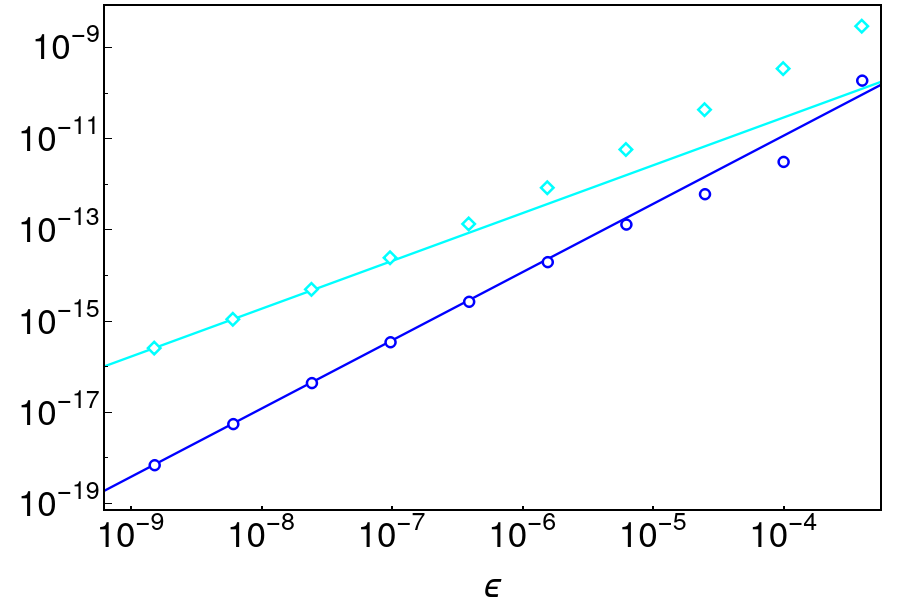} 
     \caption{Double-well model.  Log-log plots of the error in the ground state $\langle q\sigma_z \rangle$ in APT(1) (blue circles) and LW(1) (cyan diamonds) approximations versus $\e$.  Estimated asymptotic linear behaviors are $-11.36 + 1.50 \ln\e$ for APT(1) and $-14.58 + 1.05 \ln\e$ and LW(1). The coupling parameter is $\gamma=0.02$. \label{fig:gs:qz}}
\end{center}
\end{figure}

Results for the observable $\langle q^2\rangle$ and the joint electron-nucleus observable $\langle q\sigma_z \rangle$, where $\sigma_z$ is the Pauli matrix in the electronic basis $\{|1\rangle,|2\rangle\}$, are shown in Figs.~\ref{fig:gs:qSq} and \ref{fig:gs:qz}.  The calculation of 
$\langle q\sigma_z\rangle$ requires the evaluation of the generator $G_1$.  In APT, its off-diagonal matrix element is 
\begin{align}
G_{1,mn}^{APT} &= -i \f{\theta_q}{2(E_m-E_n)} \p_q -i \f{\theta_{qq}}{4(E_m-E_n)} -i \f{\theta_q \p_q E_n}{2(E_m-E_n)^2} {,}
\end{align}
while in the LW approach it is
\begin{align}
G_{1,mn}^{LW} &= -i \f{\theta_q}{2(E_m-E_n)} \p_q {.}
\end{align}

\begin{figure}[h]
\begin{center}
     \includegraphics[width=0.50\textwidth]{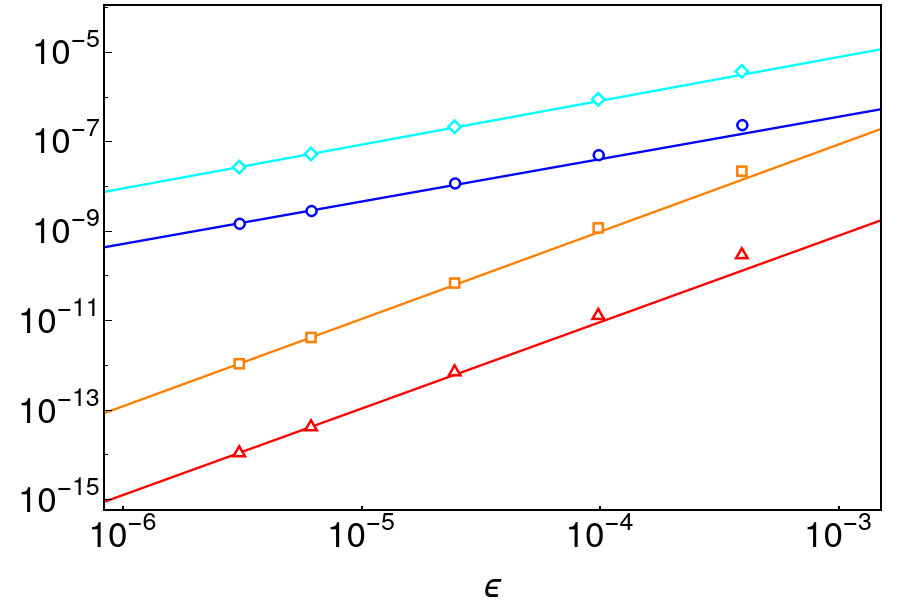} 
     \caption{Double-well model.  Log-log plot of the error in the energy of the excited state with energy closest to $-0.46$ a.u.~in APT(1) and APT(2) (blue circles and red triangles) and LW(1) and LW(2) (cyan diamonds and orange squares) approximations versus $\e$.  Estimated asymptotic linear behaviors are $-8.28 + 0.95 \ln\e$ and $-7.59 + 1.93 \ln\e$ for APT(1) and APT(2) and $-4.98 + 0.98 \ln\e$ and $-2.77 + 1.95 \ln\e$ for LW(1) and LW(2). The coupling parameter is $\gamma=0.20$. \label{fig:constant:energy}}
\end{center}
\end{figure}

\begin{figure}[h]
\begin{center}
     \includegraphics[width=0.50\textwidth]{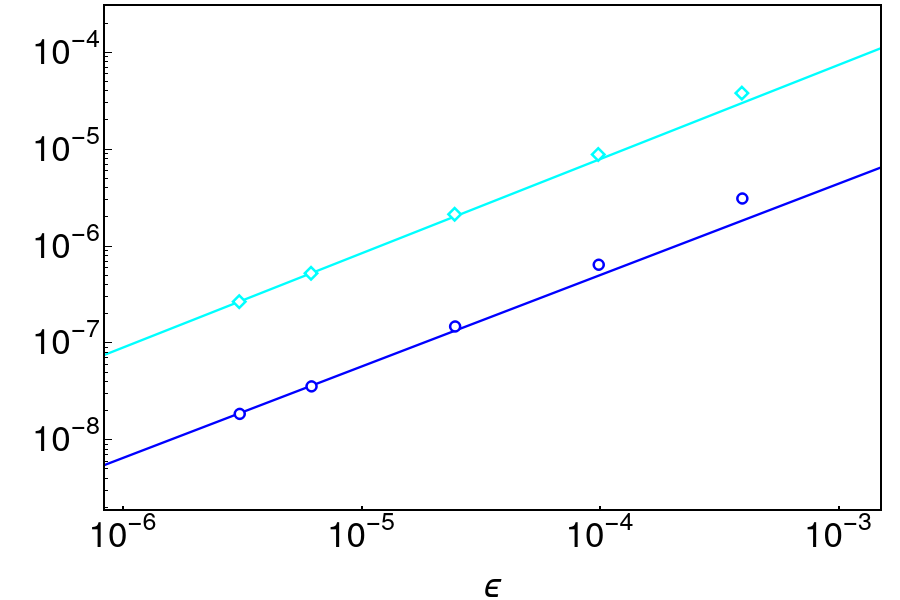} 
     \caption{Double-well model.  Log-log plot of the error in $\langle q^2 \rangle$ for the excited state with energy closest to $-0.46$ a.u.~in APT(1) (blue circles) and LW(1) (cyan diamonds) approximations versus $\e$.  Estimated asymptotic linear behaviors are $-5.81 + 0.94\ln\e$ for APT(1) and $-2.79 + 0.97 \ln\e$ for LW(1). The coupling parameter is $\gamma=0.20$. \label{fig:constant:qSq}}
\end{center}
\end{figure}

\begin{figure}[h]
\begin{center}
     \includegraphics[width=0.50\textwidth]{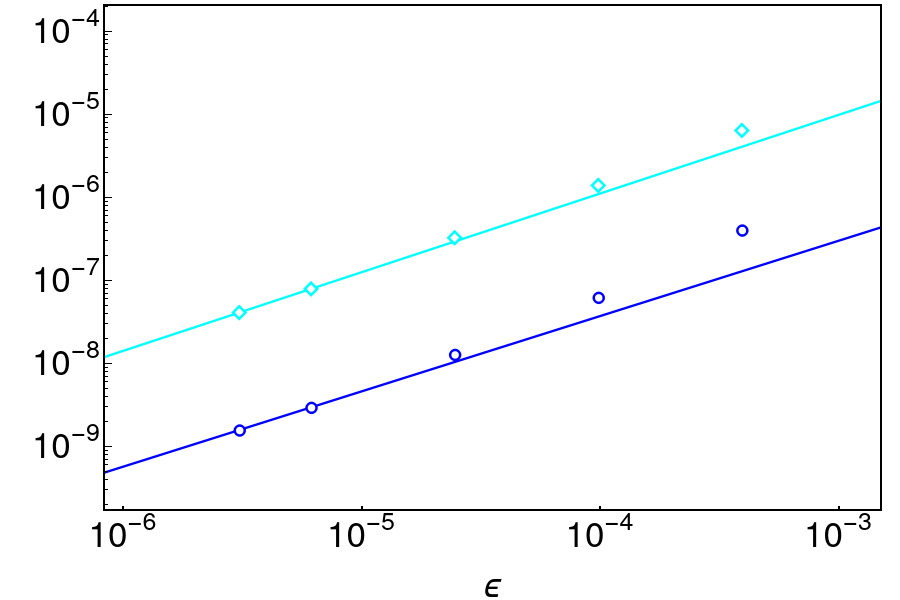} 
     \caption{Double-well model.  Log-log plot of the error in $\langle q\sigma_z \rangle$ for the excited state with energy closest to $-0.46$ a.u.~in APT(1) (blue circles) and LW(1) (cyan diamonds) approximations versus $\e$.  Estimated asymptotic linear behaviors are $-8.75 + 0.91 \ln\e$ for APT(1) and $-4.98 + 0.95 \ln\e$ for LW(1). The coupling parameter is $\gamma=0.20$. \label{fig:constant:qz}}
\end{center}
\end{figure}

Next, we make a comparison between APT and LW's theory under the condition that the energy is held constant while taking the $\e\rightarrow 0$ limit.  Since the expectation value of all energy terms other the nuclear kinetic energy are so weakly dependent on $\e$, this gives the same results as 
holding the nuclear kinetic energy constant,
which is the assumption made in Refs. \cite{littlejohn1993,weigert1993,teufel2003,matyus2019,littlejohn2024}.
In the double-well model, this corresponds to a vibrational quantum number $\n$ that scales as
$\n \sim \e^{-1/2}$, which means the nuclear wave function $\z_{n\n}$ is semiclassical.  
In this limit, the error in $\lambda$, $\langle q^2\rangle$, and $\langle q\sigma_z\rangle$ in the APT($p$) approximation has the same asymptotic scaling (to within the numerical limitations of the estimate) as the error in the LW($p$) approximation, as shown in Figs.~\ref{fig:constant:energy}-\ref{fig:constant:qz}.
%
%

We note that Born and Oppenheimer's power series expansion method \cite{born1927} fails qualitatively in the double-well model.  If the expansion in powers of $\e^{1/4}(q-q_0)$ is made with respect to the minimum of the left well, then 
it will not be practical
to capture the wave function amplitude in the right well.
Consequently, $\langle q\rangle$ will be qualitatively incorrect and the tunnel splitting, which depends nonperturbatively on $\e$, cannot be calculated.  It is also impractical to describe the wave function amplitude in the left and right wells by expanding with respect to $q=0$.

\bibliography{apt-static-2023}